\documentclass[amsmath,amssymb,aps,prb,twocolumn, superscriptaddress,longbibliography]{revtex4-1}
\usepackage{graphicx}
\usepackage{dcolumn}
\usepackage{color}
\usepackage{hyperref}

\renewcommand{\cite}[1]{[\onlinecite{#1}]}
\newcommand{\cra}[1]{\hat{a}^{\dag}_{#1}}  
\newcommand{\ana}[1]{\hat{a}_{#1}}         
\newcommand{\ket}[1]{\left|#1\right>}      

\newcommand{\eps}{\varepsilon}      
\newcommand{\df}[2]{\frac{\partial #1}{\partial #2}} 

\begin{document}

\title{Topological states in qubit arrays induced by density-dependent coupling}

\author{Andrei~A.~Stepanenko}
\affiliation{Department of Physics and Engineering, ITMO University, Saint Petersburg 197101, Russia}

\author{Mark~D.~Lyubarov}
\affiliation{Department of Physics and Engineering, ITMO University, Saint Petersburg 197101, Russia}

\author{Maxim~A.~Gorlach}
\affiliation{Department of Physics and Engineering, ITMO University, Saint Petersburg 197101, Russia}
\email{m.gorlach@metalab.ifmo.ru}


\begin{abstract}
Topological states of light open exciting possibilities in quantum photonics promising the topological protection of quantum entanglement. Here, we put forward an approach to realize the topological states of photon pairs mediated by the effective density-dependent coupling which is manifested as the dependence of the tunneling amplitude on the number of photons. As a specific platform, we investigate the arrays of nearest-neighbor coupled transmon qubits, where the effective density-dependent coupling is engineered by inserting auxiliary frequency-detuned resonators. We prove the topological origin of the designed model by the direct evaluation of the Zak phase highlighting the feasibility of our proposal for  state-of-the-art fabrication technologies.

\end{abstract}

\maketitle

\section{INTRODUCTION}

Recent developments in nanophotonics and nanofabrication technologies have enabled a class of structures utilizing the concept of topological protection on a chip~\cite{Hafezi2013,Mittal-Hafezi-2018,Blanco-Science,Blanco-Nanophotonics} paving the way towards disorder-robust quantum-optical  circuitry~\cite{Tambasco-Peruzzo,Wang-Jin-Advmat,Wang-Jin-Optica,Wang-Jin-2019,Blanco-2020}. Further progress in this direction is largely stimulated by the simultaneous advances in quantum technologies~\cite{Arute2019} which may open unprecedented possibilities in quantum simulations~\cite{McArdle2020}. In this context, it is especially important to bridge the gap between the perspective designs of quantum-photonic topological structures proposed theoretically and the designs feasible for state-of-the-art fabrication technologies.

One of the most promising experimental platforms to manipulate photonic states with non-classical statistics is currently provided by the arrays of superconducting qubits~\cite{Oliver2020}, transmon qubits~\cite{Koch2007} being a popular choice. During the recent years, the arrays of transmon qubits have also proved to be a promising platform to probe the topological physics~\cite{Roushan2016,Roushan2017,Cai2019,Besedin}. At the same time, interacting multi-photon topological phases and the associated topological transitions in qubit arrays remain vastly unexplored, which does not allow to fully harness their potential in the disorder-robust quantum circuitry.

To reveal the unique features of few-photon topological states in qubut arrays, we put forward a model featuring an interaction-induced topological state of photon pair facilitated by the  density-dependent coupling captured by the term $-T\,\hat{a}^\dag(\hat{n}_a+\hat{n}_b)\,\hat{b}+\text{H.c.}$ in the Hamiltonian, where $\hat{n}_a$ and $\hat{n}_b$ are the number of photons in qubits $a$ and $b$, respectively, while $\hat{a}$ and $\hat{b}$ are the annihilation operators for the respective qubits. 

As we demonstrate, the required density-dependent coupling can be readily engineered in the array of superconducting transmon qubits by inserting frequency-detuned auxiliary resonators. While in the previous works density-dependent coupling was treated as a parasitic side effect~\cite{Kounalakis2018}, we reveal in this Article that it may become a crucial ingredient to achieve an interaction-induced topological state of two microwave photons in an experimentally feasible geometry of qubit array.

The rest of the paper is organized as follows. In Sec.~\ref{sec:3q} we discuss the realization of density-dependent coupling due to the inserted auxiliary frequency-detuned resonators in the general multi-photon case. Using this as an elementary building block, in Sec.~\ref{sec:top} we design the system supporting the topological edge state of photon pair arising due to the effective density-dependent coupling. In Sec.~\ref{sec:num}, we simulate the proposed system numerically,  reconstruct the profile of the two-photon edge-localized state and prove the topological origin of the predicted edge state, concluding by the discussion of our results and an outlook in Sec.~\ref{sec:con}. Further details are provided in Appendixes~\ref{supp:a} and \ref{supp:b} which summarize the derivation of the effective Hamiltonian for our model and the calculation of the Zak phase.

\section{Engineering density-dependent coupling}
\label{sec:3q}

\begin{figure}[h]
    \centering
    \includegraphics[width=\linewidth]{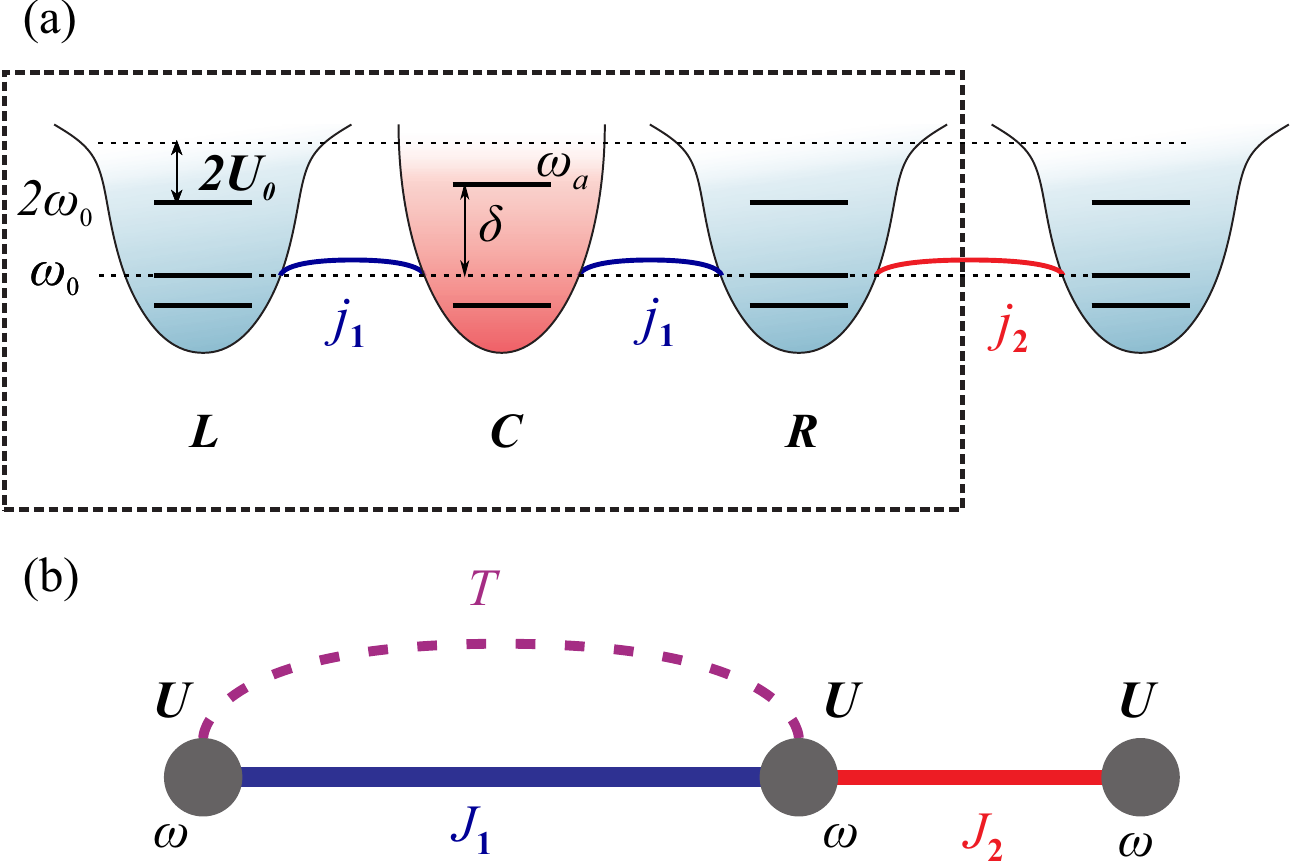}
    \caption{
    (a) Unit cell of the studied system. Black dashed rectangle illustrates the realization of the effective density-dependent coupling due to the auxiliary resonator. (b) In the limit $|\delta|\gg |j_{1,2}|$ and $|\delta|\gg |U_0|$ the system in panel (a) reduces to the effective model where all sites have the same eigenfrequencies $\omega$, anharmonicity $U$, linear couplings $J_1$ and $J_2$ as well as density-dependent coupling $T$, defined in the text.}
    \label{fig:scheme_and_mapping}
\end{figure}

First we revisit the engineering of the effective density-dependent coupling due to the inserted additional resonator performing the analysis analogously to Refs.~\cite{Jin2013,Kounalakis2018,Collodo2019}. The basic building block of our structure depicted in Fig.~\ref{fig:scheme_and_mapping}(a) in dashed rectangle includes two qubits tuned to the same frequency $\omega_0$  coupled via the auxiliary resonator with eigenfrequency $\omega_a$. Effective photon-photon interactions occur due to the anharmonicity $U_0$ of qubits [Fig.~\ref{fig:scheme_and_mapping}(a)], while the auxiliary resonators are strictly harmonic and therefore can be realized just as LC resonators. Qubit-resonator and qubit-qubit coupling constants are equal to $j_1$ and $j_2$, respectively and the overall system is described by the Bose-Hubbard Hamiltonian:
\begin{eqnarray}
\label{eq:FullHamiltonian}
\hat{H} & =& \hat{H}_0+\hat{V}\:,\\
\label{eq:Hamiltonian}
  \hat{H_0} &=& \omega_0(\hat{n}_R+\hat{n}_L)+\omega_a\hat{n}_C\nonumber\\ &&+ U_0\hat{n}_R(\hat{n}_R-1) + U_0\hat{n}_L(\hat{n}_L-1)\:,\\
  \label{eq:Perturb}
\hat{V} &=&  - j_1 (\cra{L} \ana{C}+\cra{C} \ana{L}) -  j_1 (\cra{R} \ana{C}+\cra{C} \ana{R})\:,
\end{eqnarray}
where we set $\hbar=1$ for simplicity.

We note that such system conserves the total number of photons since $[\hat{H},\hat{n}_R+\hat{n}_C+\hat{n}_L]=0$, and, therefore, each eigenstate has a well-defined total number of photons $N$.

For the given $N$, the energies of the eigenstates split into several distinct groups having different energy scales: $\eps\approx N\,\omega_0$, $\eps\approx (N-1)\,\omega_0+\omega_a$, $\dots$, $\eps\approx N\,\omega_a$, where the splitting between these groups is proportional to the detuning $\delta=\omega_a-\omega_0$. In our analysis, we are interested in the first group of eigenstates when photons  predominantly localize in qubits. For the remaining states, a sizeable part of photons is localized in the auxiliary resonators and thus the effects of interaction are less pronounced.

Assuming that the frequency difference $\delta=\omega_a-\omega_0$ is much larger than the other parameters of the system such as $U_0$, $j_1$ and $j_2$, we derive the effective Hamiltonian for the states $\eps\approx N\,\omega_0$. To exclude the remaining states, we apply the degenerate second-order perturbation theory~\cite{Bir} as further discussed in Appendix~\ref{supp:a}. In this approximation, the only nonzero off-diagonal matrix elements of the effective Hamiltonian are between the states $\ket{N_L+1,0,N_R}$ and $\ket{N_L,0,N_R+1}$, whereas the effective coupling between the states $\ket{N_L+s,0,N_R}$ and $\ket{N_L,0,N_R+s}$, $s\geq 2$ appears to be negligible.

The resulting effective Hamiltonian for a pair of qubits coupled via the auxiliary resonator reads:
\begin{eqnarray}
\hat{H}^{(\rm{eff})}&=&\left(
\begin{array}{cc}
   H_{LL} & H_{LR} \\
   H_{RL} &  H_{RR}\\
\end{array}
\right)\:,\label{eq:33q}\\
H_{LL}& =& \omega_0\,N + U_0 (N_L+1) N_L\nonumber\\&&+ U_0 N_R(N_R-1) - \dfrac{j_1^2 (N_L+1)}{\delta-2U_0 N_L}\:,\label{eq:HLL}\\
H_{RR}&=& \omega_0\,N + U_0 N_L(N_L-1)\nonumber\\&&+ U_0 (N_R+1) N_R - \dfrac{j_1^2 (N_R+1)}{\delta-2U_0 N_R}\:,\label{eq:HRR}
\end{eqnarray}

\begin{eqnarray}
\label{eq:HLR}
H_{LR}&=&H_{RL}=-J^{(\rm{eff})}\,\sqrt{(N_L+1)(N_R+1)}\:,\\
\label{eq:Jeff}
J^{(\rm{eff})}&=&\dfrac{j_1^2}{2}\left(  \dfrac{1}{\delta-2U_0N_L}+\dfrac{1}{\delta-2U_0N_R}\right)\:,
\end{eqnarray}
where we exploit the basis composed of $\ket{N_L+1,0,N_R}$ and $\ket{N_L,0,N_R+1}$ vectors so that the total number of photons $N=N_L+N_R+1$ and the intermediate state $\ket{N_L,1,N_R}$ is excluded. Equations~\eqref{eq:33q}-\eqref{eq:Jeff} suggest that the sites labelled by $L$ and $R$  acquire an additional $N$-dependent correction to their energy, which is an additional source of anharmonicity along with $U_0$ term. Moreover, the effective coupling constant $J^{(\rm{eff})}$ also becomes the function of $N$, i.e., the coupling between qubits is density-dependent.

To separate the linear coupling from the density-dependent part, we consider the case $N_L=N_R=0$ that corresponds to the single photon travelling from the left qubit to the right. The coupling constant in such case reads
\begin{equation}\label{eq:LinearCoupling}
J_1=\frac{j_1^2}{\delta}   
\end{equation}
which corresponds to the usual linear coupling constant. The remaining part
\begin{equation}
\begin{split}
J^{(\rm{eff})}-J_1=\frac{j_1^2\,U_0}{\delta}\,\left(\frac{N_L}{\delta-2U_0\,N_L}+\frac{N_R}{\delta-2U_0\,N_R}\right)\\
\approx \frac{j_1^2\,U_0}{\delta^2}\,(N-1)\,
\end{split}
\end{equation}
explicitly depends on the total number of photons being associated with the density-dependent coupling.

In our analysis, we investigate the case of two photons, which is the minimal number of particles needed to observe interaction-induced effects. To provide a simple description of the designed qubit array in the two-photon regime, we introduce a  Bose-Hubbard type effective Hamiltonian with a separate term $\propto T$ responsible for density-dependent coupling. For the pair of qubits it reads:
\begin{gather}
\hat{H}_{2q}^{(\rm{eff})}=\omega\,(\hat{n}_1+\hat{n}_2)+U\,\left[\hat{n}_1\,(\hat{n}_1-1)+\hat{n}_2\,(\hat{n}_2-1)\right]\notag\\
-J_1\,\left(\cra{1}\ana{2}+\text{H.c.}\right)-T\,\left[\cra{1}\,(\hat{n}_1+\hat{n}_2)\,\ana{2}+\text{H.c.}\right]\:,\label{eq:TwoQubitEffH}
\end{gather}
where $\omega$ and $U$ are renormalized eigenfrequency and anharmonicity of the individual qubit, $J_1$ is the effective single-photon hopping rate, while $T$ describes the strength of density-dependent coupling. In the basis of the two-photon states $\ket{2,0}$, $\ket{1,1}$ and $\ket{0,2}$ this Hamiltonian yields $3\times 3$ matrix
\begin{equation}\label{eq:TwoQubitH}
\hat{H}_{2q}^{(\rm{eff})}=\begin{pmatrix}
2\omega+2U & -\sqrt{2}\,(J_1+T) & 0\\
-\sqrt{2}\,(J_1+T) & 2\omega & -\sqrt{2}\,(J_1+T)\\
0 & -\sqrt{2}\,(J_1+T) & 2\omega+2U
\end{pmatrix}\:.
\end{equation}
On the other hand, the matrix of the same form can be obtained from $6\times 6$ Hamiltonian for the states $\ket{2,0,0}$, $\ket{1,1,0}$, $\ket{1,0,1}$, $\ket{0,2,0}$, $\ket{0,1,1}$, $\ket{0,0,2}$ excluding the states $\ket{1,1,0}$, $\ket{0,2,0}$ and $\ket{0,1,1}$ via perturbation theory (Appendix~\ref{supp:a}). This provides the identification of the effective Hamiltonian parameters:
\begin{gather}
    \omega=\omega_0-\frac{j_1^2}{\delta}\:,\label{eq:OmegaEff}\\
    U=U_0-\frac{2j_1^2\,U_0}{\delta\,(\delta-2U_0)}\:,\\
    T=\frac{j_1^2\,U_0}{\delta\,(\delta-2U_0)}\:,\label{eq:TEff}
\end{gather}
where $J_1$ is defined by Eq.~\eqref{eq:LinearCoupling}.

\section{Topological states in the model with density-dependent coupling}
\label{sec:top}

\begin{figure}[h!]
    \centering
    \includegraphics[width=0.68\linewidth]{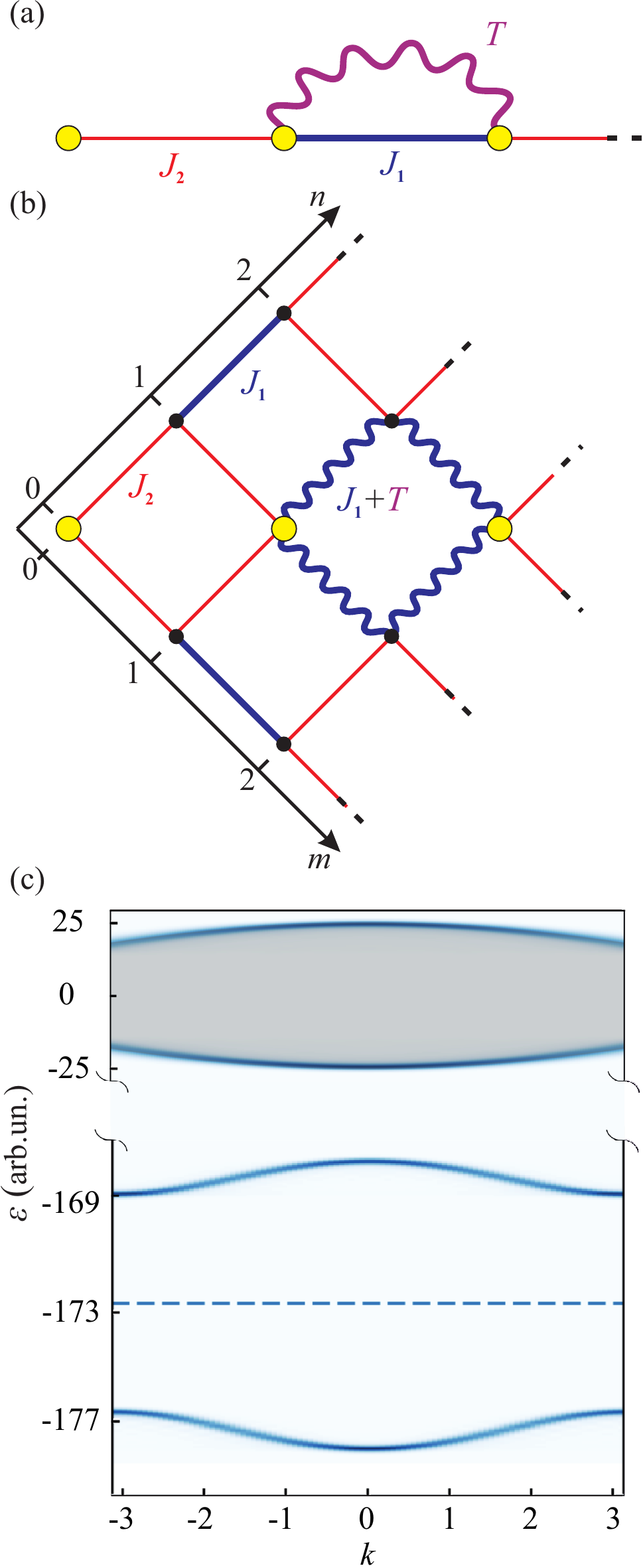}
    \caption{
    Mapping of quantum one-dimensional two-photon problem onto the classical two-dimensional system. (a) Two-particle problem with single-particle tunneling constants $J_1$ and $J_2$ (straight blue and red lines) and density-dependent coupling $T$ (wavy purple line). Yellow circles indicate on-site anaharmonicity $U$. (b) The respective 2D tight-binding model. Coordinates $m$ and $n$ correspond to the positions of the first and second photons in the original problem. Sites with $m=n$ experience energy shift $2U$. Wavy lines demonstrate the modification of the tunneling links related to the density-dependent coupling in the original problem. (c) Calculated dispersion of the two-photon states for the Hamiltonian Eq.~\eqref{eq:Hamiltonian_T} with parameters $J_1 = 5;\, J_2 = 7.5;\, T = -25;\, U = -84;\, U_1=-2.27$. Two lower bands with the energies in the range $-180<\eps<-165$ correspond to bulk doublons, whereas higher-energy states with $-25<\eps<25$ form the scattering continuum. Dashed line shows the energy of the edge state, $\eps_{\rm{edge}} \simeq -173$. $2\omega$ is used as an energy reference.
    }
    \label{fig:2D_scheme}
\end{figure}

As a next step, we harness the mechanism of density-dependent coupling designing an interaction-induced topological state. To this end, we consider a periodic structure with the unit cell shown in Fig.~\ref{fig:scheme_and_mapping}(a) and containing an alternating pattern of couplings one of which is density-dependent, whereas the second one is the standard linear coupling. The effective Hamiltonian of such system is a straightforward generalization of Eq.~\eqref{eq:TwoQubitEffH}:
\begin{eqnarray}
\label{eq:Hamiltonian_T}
  \hat{H} &=& \omega\sum_{m}\hat{n}_m+ U\sum_{m}\hat{n}_m(\hat{n}_m-1)+\delta U_1\,\hat{n}_1(\hat{n}_1-1)\nonumber\\
  && -  J_1 \sum_{m}(\cra{2m} \ana{2m-1}+\cra{2m-1} \ana{2m})\nonumber\\
  &&- J_2 \sum_{m}(\cra{2m} \ana{2m+1}+\cra{2m+1} \ana{2m})
  \nonumber\\
  &&-T\sum_m (\cra{2m}(\hat{n}_{2m}+\hat{n}_{2m-1})\ana{2m-1}+\text{H.c.})\:,
\end{eqnarray}
where $J_2$ is the coupling constant of the adjacent qubits and the remaining parameters are defined above. We also introduce an additional correction $\delta U_1$ to the anharmonicity of the first qubit to compensate the lack of neighbors as discussed below. The proposed system is depicted schematically in Fig.~\ref{fig:2D_scheme}(a).

The two-photon eigenstates are found as the solutions of the Schr\"odinger equation $\hat{H}\ket{\psi} = (\eps+2\omega)\,\ket{\psi}$ with the Hamiltonian \eqref{eq:Hamiltonian_T} and the wave function 
\begin{equation}\label{eq:WaveFunc}
\ket{\psi} = \frac{1}{\sqrt{2}}\,\sum_{m,n} b_{mn}\,\cra{m}\cra{n}\ket{0}\:,
\end{equation}
where $b_{mn}=b_{nm}$ due to the bosonic nature of the problem. This yields the linear system of equations for $b_{mn}$ coefficients which can be reinterpreted as a two-dimensional single-particle tight-binding problem illustrated in Fig.~\ref{fig:2D_scheme}(b).

To grasp the main features of this model, we take the limit of strong anharmonicity $U$, in which case Bose-Hubbard model predicts the emergence of bound boson pairs (doublons)~\cite{Mattis1986,Winkler2006,Valiente2008,Liberto2016,Bello2017} spectrally isolated from the rest of the two-particle states. Note that the signatures of such two-photon bound states have recently been observed experimentally in the arrays of superconducting qubits as well~\cite{Yan-Pan-2019,Ye-Pan-2019,Besedin}.

Tight co-localization of photons provides a key to doublon analytical description: we keep the dominant terms $b_{nn}$ and $b_{n,n\pm 1}$ in the doublon wave function Eq.~\eqref{eq:WaveFunc} neglecting all terms $b_{mn}$ with $|m-n|\geq 2$. This yields the system
\begin{eqnarray}
\left(\eps -2(U+\delta U_1)\right)\,b_{00} &=& 2J_2\,b_{10},\\
\eps\,b_{10} &=& J_2\,b_{00} + J_2\,b_{11},\label{eq:b10}\\
\left(\eps -2U\right)b_{11} &=& 2J_2\,b_{10} + 2\,(J_1+T)\, b_{21}\:,\\
\eps\,b_{21} &=& (J_1+T) (b_{11}+b_{22})\label{eq:b21},\\
&...&\nonumber
\end{eqnarray}
To simplify the analysis further, we treat $b_{n,n-1}$ coefficients as a perturbation and express them from Eqs.~\eqref{eq:b10}, \eqref{eq:b21} setting $\eps\approx 2U$, which is the zeroth order approximation for the doublon energy. Such procedure yields  approximate equations for $\beta_{nn}$ coefficients with $1/U$ precision:
\begin{eqnarray}
\left[\eps -2(U+\delta U_1)-\dfrac{J_2^2}{U}\right]b_{00} &=& \dfrac{J_2^2}{U} b_{11},\nonumber\\
\left[\eps -2U-\dfrac{J_2^2}{U}-\dfrac{(J_1+T)^2}{U}\right]b_{11} &=& \dfrac{J_2^2}{U} b_{00} + \dfrac{(J_1+T)^2}{U} b_{22},\nonumber\\
\left[\eps -2U-\dfrac{J_2^2}{U}-\dfrac{(J_1+T)^2}{U}\right]b_{22} &=& \dfrac{J_2^2}{U} b_{33} + \dfrac{(J_1+T)^2}{U} b_{11}.\nonumber\\
&...&\label{eq:SSH-model}
\end{eqnarray}
Equations~\eqref{eq:SSH-model} correspond to the well-celebrated Su-Schrieffer-Heeger model with the alternating tunneling links $J_2^2/U$ and $(J_1+T)^2/U$. Additionally, if we choose anharmonicity detuning of the edge qubit $\delta U_1=(J_1+T)^2/(2U)$, the frequency of the edge site in the effective model will not be detuned which guarantees that the topological state appears in the middle of bandgap with the localization at the weak link edge.

To confirm our expectation, we retrieve the dispersion of bound pairs from numerical simulations. For that purpose, we simulate a finite array of 29 sites with parameters $J_1 = 5;\, J_2 = 7.5;\, T = -25;\, U = -84;\, \delta U_1=-2.27$ and extract energies $E_{\alpha}$ and the wave functions $\ket{\psi^{(\alpha)}}$ of extended doublon states. Fourier-transforming each of the wave functions, we obtain the set of amplitudes
\begin{equation}
\tilde{\psi}^{(\alpha)}(k)=\sum\limits_{m,n}\,b_{mn}^{(\alpha)}\,\exp\left[-ik\,(m+n)/4\right]\:,
\end{equation}
where $k$ is swapped through the entire first Brillouin zone from $-\pi$ to $\pi$. Finally, we evaluate the density of states
\begin{equation}
F(\eps,k)=\sum\limits_{\alpha}\,\left|\tilde{\psi}^{(\alpha)}(k)\right|^2\,\exp\left(-\frac{(\eps-E_{\alpha})^2}{2\,\sigma^2}\right)
\end{equation}
which is plotted in Fig.~\ref{fig:2D_scheme}(c). Since the maxima of the function $F(\eps,k)$ are achieved for $\eps\approx E_{\alpha}$, Fig.~\ref{fig:2D_scheme}(c) illustrates the dispersion of bound pairs provided the number of the sites chosen for numerical simulations is large enough.

Comparing the calculated energies of the eigenstates $E_{\alpha}$ with the dispersion of bulk two-photon states, we recover that one of the states with energy $\eps_{\rm{edge}}\approx -173$ arises in the middle of doublon bandgap being localized at the edge with $J_2$ link. Note that this result is in contrast with the single-particle case, when the topological state is formed on the opposite edge of the array terminated by $J_1$ link.

Hence, we conclude that the predicted two-photon edge state has an interaction-induced nature, being facilitated by the density-dependent coupling.

\section{Simulation of the full model}
\label{sec:num}

\begin{figure}[t!]
    \centering
    \includegraphics[width=\linewidth]{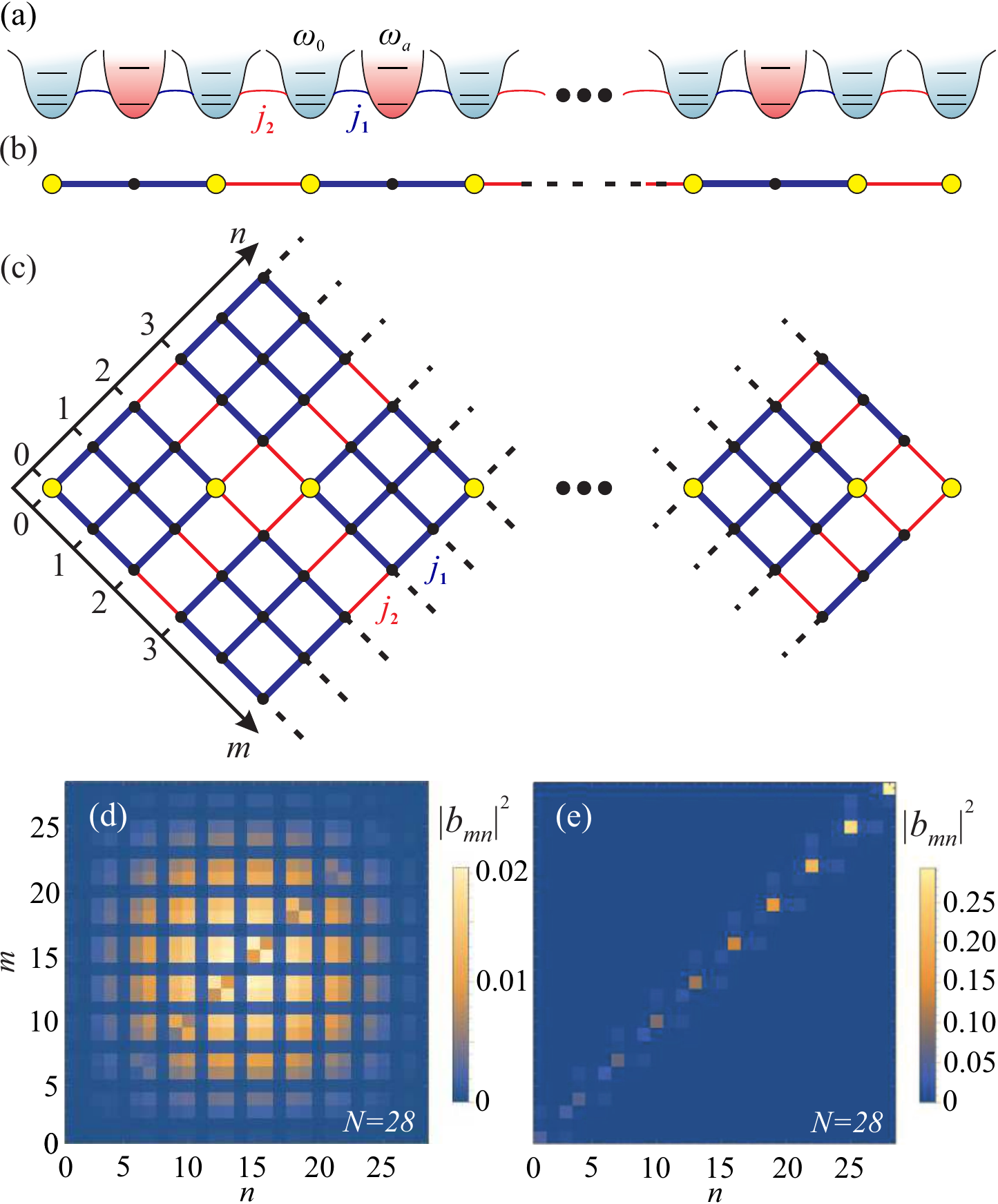}
    \caption{
    (a) Schematic of the studied array of qubits shown by gray with auxiliary resonators facilitating density dependent coupling highlighted by red. Qubits at right and left edges are connected to bulk sites via the different links $j_1$ and $j_2$ which makes the edges of the array inequivalent.  (b) Effective one-dimensional model that corresponds to (a). Yellow circles indicate on-site anharmonicity $U$ and black additional site inserted between between them corresponds to the frequency-detuned resonator. (c) Mapping of the original system onto 2D tight-binding model.  (d) The probability distribution for the two-photon scattering state ($\eps/j_2 \sim 2\omega_0/j_2 = 275.9$) obtained by numerical simulation of the system of 28 qubits with $\delta/j_2 = -34.5; j_1/j_2 = 5.86; U_0/j_2 = 0$. The effective density-dependent coupling $T/j_2=0$, no two-photon edge states are observed. (e) Two-photon probability distribution for the doublon edge state ($\eps/j_2 = 270.7$) in the array of 28 qubits with $U_0/j_2=-3.5$, which yields $T/j_2\sim -0.14$. Localization at $j_2$ link is observed. }
    \label{fig:edge_states}
\end{figure}

Having predicted the interaction-induced two-photon edge state in the simplified model Eq.~\eqref{eq:Hamiltonian_T}, we now turn to the full model [Fig.~\ref{fig:edge_states}(a)] explicitly simulating the auxiliary resonators used to engineer the effective density-dependent coupling [Fig.~\ref{fig:edge_states}(b)].

Similarly to the analysis in Sec.~\ref{sec:top}, we map the original two-particle problem onto the equivalent 2D tight-binding setup shown in Fig.~\ref{fig:edge_states}(c) and calculate its eigenstates numerically for the parameters $\omega_0/j_2=138.0$, $\delta/j_2 = -34.5, j_1/j_2 = 5.86$ giving rise to quite similar physics as in Sec.~\ref{sec:top}.  Note that the respective single-photon model appears to be trivial since two effective coupling constants, $J_1=j_1^2/\delta$ and $j_2$ are equal to each other with high precision.


First we examine the case when on-site anharmonicity of qubits $U_0=0$ and the effective density-dependent coupling vanishes [cf. Eq.~\eqref{eq:TEff}]. As a result, bound photon pairs are absent, and the only possible type of the two-photon states is the scattering state depicted in Fig.~\ref{fig:edge_states}(d) showing the absence of photon co-localization. Furthermore, no edge-localized two-photon states are observed.

However, switching on the effects of interaction for $U_0/j_2=-3.5$, we observe the formation of bulk doublons and the two-photon edge state at $j_2$ link [Fig.~\ref{fig:edge_states}(e)] as has been anticipated from our simplified model in Sec.~\ref{sec:top}. Checking the behavior of $b_{nn}$ coefficients, we observe that they are negligibly small in the auxiliary resonators featuring an exponential decay with $n$ for qubits. Therefore, the obtained result provides a clear evidence of the interaction-induced two-photon edge state.

Next we probe the topological origin of the predicted edge state extracting the Zak phase for bulk doublon bands. Note that the direct calculation of the Zak phase based on Berry connection evaluation in the entire Brillouin zone is quite complicated task which requires an analytical solution for the dispersion of bulk doublons~\cite{Gorlach2017,Stepanenko2020}. An alternative less computationally expensive approach is to assess the behavior of the wave function only in high-symmetry points of the Brillouin zone and retrieve the Zak phase as
\begin{equation}
    \gamma = \alpha(0)-\alpha(\pi)\:,
\end{equation}
where $\alpha(k)$ is the phase acquired by the Bloch mode with wave number $k$ under the inversion (see further details in Appendix~\ref{supp:b}). The eigenmodes of a finite array are the superpositions of Bloch modes with wave numbers $k$ and $-k$. Two special points in the Brillouin zone, $0$ and $\pi$, which are of interest for us, are invariant under reflection. Therefore, we can extract the required phases from the numerical simulation of a finite array.

\begin{figure}[ht!]
    \centering
    \includegraphics[width=\linewidth]{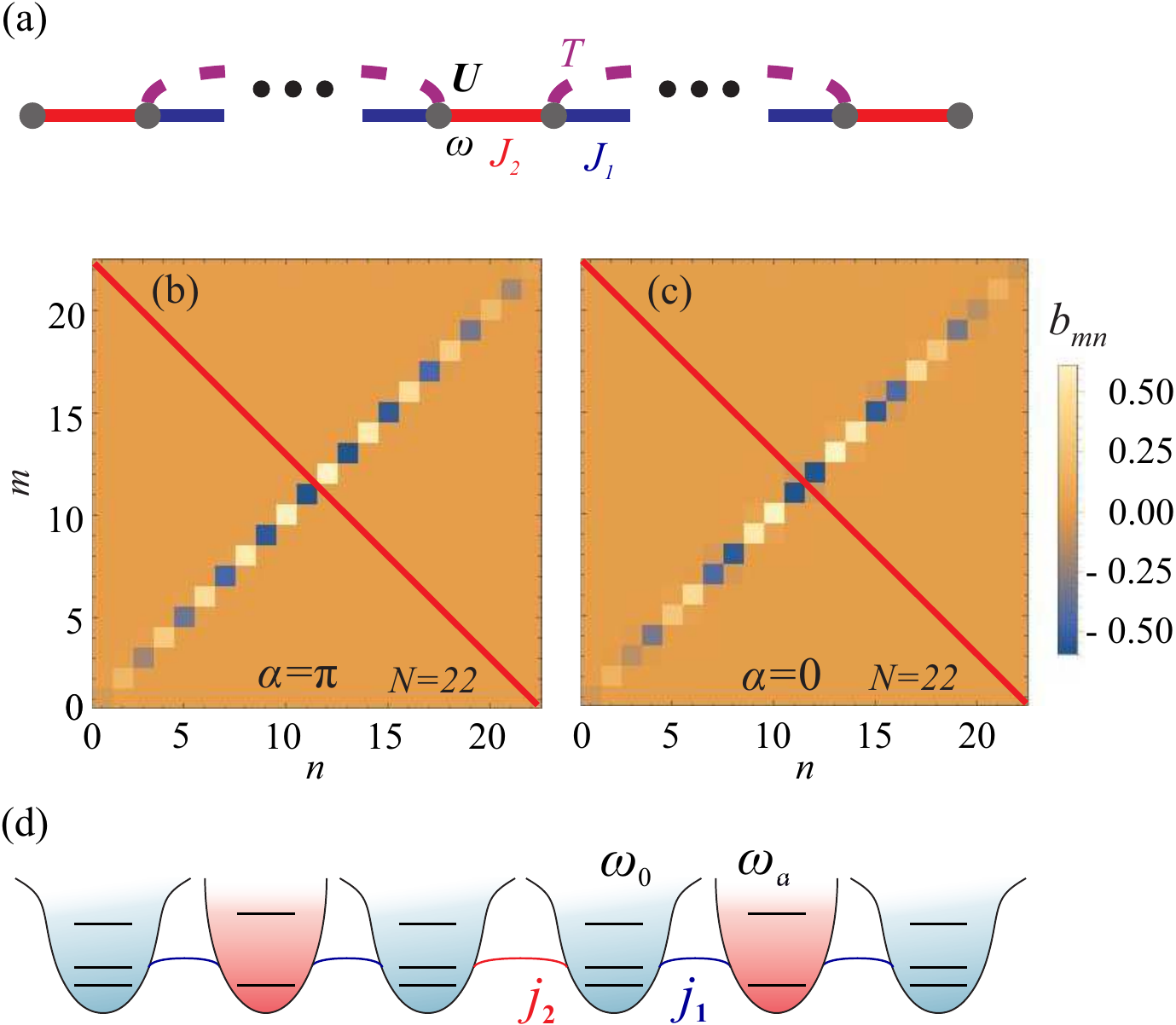}
    \caption{(a) The scheme of qubit array with density-dependent coupling $T$. The center of  inversion is located at $J_2$ link. (b,c) Superposition coefficients $b_{mn}$ for bulk doublon state with $k=0$ ($\eps\simeq -168.0$) and $k=\pi$ ($\eps\simeq -169.3$), respectively. The red line shows the symmetry axis. The wave function is anti-symmetric (symmetric), so $\alpha(0) = \pi$ and $\alpha(\pi)=0$. (d) Unit cell of the system with effective density-dependent coupling realized via the auxiliary detuned resonator. The tunneling link $j_2$ is located in the middle of the array.} 
    \label{fig:zak_phase}
\end{figure}

The Zak phase intrinsically depends on the unit cell choice. Since we investigate the edge state localized at $j_2$ link, the center of inversion should be chosen to coincide with the center of this link as illustrated in Fig.~\ref{fig:zak_phase}(a) and the overall array should be inversion-symmetric. 

Using this technique, we first examine a finite array of $N=22$ qubits described by the simplified model Eq.~\eqref{eq:Hamiltonian_T} with the same parameters as in Sec.~\ref{sec:top}. Calculating the wave functions of bulk doublon states from the upper doublon band, we recover the 2D maps of $b_{mn}$ coefficients depicted in Fig.~\ref{fig:zak_phase}(b,c). When the inversion transformation is applied, the entire map of $b_{mn}$ coefficients is reflected relative to the axis shown by red. Examining the obtained maps, we notice that the mode with $k=0$ is antisymmetric, whereas the mode with $k=\pi$ is symmetric. Hence, the Zak phase $\gamma=\pi\mspace{4mu}\text{mod}\mspace{4mu} 2\pi$, which proves the topological origin of the predicted edge state. 

The same result for the Zak phase is obtained within the full model with parameters specified in Fig.~\ref{fig:edge_states}(e) caption. Quite importantly, the proposed design can be implemented experimentally using existing transmon qubits with the typical resonance frequencies $f_0=4$~GHz, anharmonicity $U_0 = - 100$~MHz and coupling strengths $j_1=170$~MHz and $j_2=29$~MHz.

\section{DISCUSSION AND CONCLUSIONS}
\label{sec:con}
In conclusion, we have demonstrated that the two-photon topological states can be facilitated by the density-dependent coupling treated previously as a parasitic side effect. Such mechanism can be readily implemented by inserting auxiliary frequency-detuned resonators, which is feasible for existing architectures of qubit arrays.

Quite remarkably, the predicted edge state has an interaction-induced origin being absent in the single-photon case and thus highlighting the potential of few-body topological states.

We believe that our study thus bridges a gap between a series of theoretical proposals for interaction-induced topological states of entangled bosons and current fabrication capabilities of qubit arrays.

\section{ACKNOWLEDGMENTS}
This work was supported by the Russian Science Foundation (Grant No.~20-72-10065). A.A.S. acknowledges partial support by Quantum Technology Centre, Faculty of Physics, Lomonosov Moscow State University. A.A.S. and M.A.G. acknowledge partial support by the Foundation for the Advancement of Theoretical Physics and Mathematics ``Basis".

\appendix

\section{Derivation of the effective Hamiltonian with density-dependent coupling}
\label{supp:a}

In this Appendix, we discuss the derivation of the effective Hamiltonian for a pair of qubits coupled via the auxiliary frequency-detuned resonator employing the degenerate second-order perturbation theory~\cite{Bir} under the assumption $|\delta| \gg |j_{1}|$ and $|\delta|\gg |U_0|$.

First we derive the effective Hamiltonian for the pair of states $\ket{1}\equiv\ket{N_L+1,0,N_R}$ and $\ket{3}\equiv\ket{N_L,0,N_R+1}$ coupled via the intermediate state $\ket{2}\equiv\ket{N_L,1,N_R}$. Note that we do not need to take into account other intermediate states since they do not provide any second-order corrections to the effective Hamiltonian.

Applying Bose-Hubbard model Eqs.~\eqref{eq:FullHamiltonian}-\eqref{eq:Perturb}, we recover the following elements of the full $3\times 3$ Hamiltonian:
\begin{gather}
H_{11} =\omega_0\,N + U_0 (N_L+1) N_L+U_0 N_R(N_R-1),\\
H_{22} =\omega_0\,(N_L+N_R)+\omega_a + U_0 N_L (N_L-1) \nonumber \\ + U_0 N_R(N_R-1),\\
H_{33} = \omega_0\,N+ U_0 N_L(N_L-1)+ U_0 (N_R+1) N_R,\\
H_{12}=H_{21}=-j_1\sqrt{N_L+1}\\
H_{23}=H_{32}=-j_1\sqrt{N_R+1}\\
H_{13}=H_{31}=0\:,
\end{gather}
where $N=N_L+N_R+1$. Next we exclude state $\ket{2}$ using the expression~\cite{Bir}:
%
\begin{multline}
 H^{\rm{(eff)}}_{mm'}=H_{mm'}+\dfrac{1}{2}\,\sum_s H_{ms} H_{sm'}\times \\ \times\left[\dfrac{1}{H_{mm}-H_{ss}}+\dfrac{1}{H_{m'm'}-H_{ss}}\right]\:,  
\end{multline}
%
where index $s$ labels the intermediate states and off-diagonal elements of the Hamiltonian are treated as a perturbation. With such an approach, we derive:
\begin{eqnarray}
H_{LL} &=& \hat{H}^{{\rm (eff)}}_{11}=H_{11}+\dfrac{|H_{12}|^2 }{H_{11}-H_{22}}\nonumber\\
&=& \omega_0\,N + U_0 (N_L+1) N_L\nonumber\\&&+ U_0 N_R(N_R-1) - \dfrac{j_1^2 (N_L+1)}{\delta-2U_0 N_L}\:,\\
H_{RR}&=& \hat{H}^{\rm{(eff)}}_{33}=H_{33}+\dfrac{|H_{23}|^2 }{H_{33}-H_{22}}\nonumber\\
&=& \omega_0\,N + U_0 N_L(N_L-1)\nonumber\\&&+ U_0 (N_R+1) N_R - \dfrac{j_1^2 (N_R+1)}{\delta-2U_0 N_R}\:,\\
H_{LR}&=&H_{RL} = \hat{H}^{\rm{(eff)}}_{13}=\hat{H}^{\rm{(eff)}}_{31}\nonumber\\
&=&\dfrac{H_{12} H_{23}}{2}\left[\dfrac{1}{H_{11}-H_{22}}+\dfrac{1}{H_{33}-H_{22}}\right]\\
&=&-J^{(\rm{eff})}\,\sqrt{(N_L+1)(N_R+1)}\:,\\
J^{(\rm{eff})}&=&\dfrac{j_1^2}{2}\left(  \dfrac{1}{\delta-2U_0N_L}+\dfrac{1}{\delta-2U_0N_R}\right)\:.
\end{eqnarray}

To capture the two-photon physics of two qubits coupled via the auxiliary resonator, we consider the set of all possible two-photon states $\ket{2,0,0}$, $\ket{1,1,0}$, $\ket{1,0,1}$, $\ket{0,2,0}$ $\ket{0,1,1}$ and $\ket{0,0,2}$. In such basis, the Hamiltonian is presented as $6\times 6$ matrix:
\begin{small}
\begin{equation}
\begin{split}
& \hat{H}_{2q}=\\
& \begin{pmatrix}
2\omega_0+2U_0 & -j_1\,\sqrt{2} & 0 & 0 & 0 & 0\\
-j_1\,\sqrt{2} & \omega_0+\omega_a & -j_1 & -j_1\,\sqrt{2} & 0 & 0\\
0 & -j_1 & 2\omega_0 & 0 & -j_1 & 0\\
0 & -j_1\,\sqrt{2} & 0 & 2\omega_a & -j_1\,\sqrt{2} & 0\\
0 & 0 & -j_1 & -j_1\,\sqrt{2} & \omega_0+\omega_a & -j_1\,\sqrt{2}\\
0 & 0 & 0 & 0 & -j_1\,\sqrt{2} & 2\omega_0+2U_0
\end{pmatrix}
\:.
\end{split}
\end{equation}
\end{small}
Next, we exclude the states 2, 4 and 5 using the degenerate perturbation theory. This yields $3\times 3$ matrix having the same structure as Eq.~\eqref{eq:TwoQubitH}. Comparing these two versions of the same Hamiltonian, we immediately identify $\omega$, $U$ and $T$ given by Eqs.~\eqref{eq:OmegaEff}-\eqref{eq:TEff} in the main text.

\section{Evaluation of the Zak phase}
\label{supp:b}

In this section, we discuss the calculation of the Zak phase for inversion-symmetric one-dimensional system performing our analysis in close analogy with Ref.~\cite{Hughes2011}.

Consider an inversion-symmetric system with the periodic part of the wave function $\ket{u_k}$. To evaluate the Zak phase, we generally need to construct Berry connection
\begin{equation}\label{BerryConn}
A(k)=i\left<u_k\left|\df{u_k}{k}\right.\right>
\end{equation}
and then calculate the integral
\begin{equation}\label{Zak1}
\gamma=\int\limits_{-\pi}^{\pi}\,A(k)\,dk\:.
\end{equation}
For complex systems, neither explicit form of $\ket{u_k}$, nor its $k$-derivative are known which makes the calculation of the integral Eq.~\eqref{Zak1} cumbersome.

However, this calculation can be considerably simplified for the case of  inversion-symmetric system. Inversion symmetry implies that
\begin{equation}\label{Inversion}
\hat{P}\,\ket{u_k}=e^{i\alpha(k)}\,\ket{u_{-k}}\:,
\end{equation}
where $\hat{P}$ is a unitary operator describing the inversion, which does not depend on $k$, and $\alpha$ is a phase which generally depends on $k$.

Double application of inversion should yield the original vector $\ket{u_k}$. On the other hand,
\begin{equation}
\hat{P}^2\,\ket{u_k}=e^{i\alpha(k)}\,\hat{P}\,\ket{u_{-k}}=e^{i\left[\alpha(k)+\alpha(-k)\right]}\,\ket{u_k}\:.
\end{equation}
Hence,
\begin{equation}\label{PhaseBehavior}
\alpha(-k)+\alpha(k)=0 \mspace{6mu}\text{mod}\mspace{4mu} 2\pi\:.
\end{equation}
Now, making use of the property Eq.~\eqref{Inversion}, we establish the link between $A(-k)$ and $A(k)$

\begin{equation}
\ket{\df{u_{-k}}{(-k)}}=i\df{\alpha}{k}\,\ket{u_{-k}}-e^{-i\alpha(k)}\,\hat{P}\,\ket{\df{u_k}{k}}\:,
\end{equation}
and calculate
\begin{equation}\label{Ak}
\begin{split}
A(-k)\equiv i\left<u_{-k}\left|\df{u_{-k}}{(-k)}\right.\right>=-\df{\alpha}{k}-A(k)\:.
\end{split}
\end{equation}
Using Eq.~\eqref{Ak}, we rewrite the expression for the Zak phase in the form:
\begin{equation}
\gamma=-\int\limits_0^{\pi}\,\df{\alpha(k)}{k}\,dk=\alpha(0)-\alpha(\pi)\:,
\end{equation}
where phase $\alpha$ describes the behavior of the wave function under inversion. Note that in order to calculate the Zak phase one neither needs to know the behavior of $\ket{u_k}$ in the entire Brillouin zone nor its explicit expression. The only crucial ingredient is the behavior of the wave function under inversion in two time-reversal-invariant points: $k=0$ and $k=\pi$. This recipe can be easily applied to calculate the Zak phase for the variety of systems including those studied above.

\bibliography{ddh_bib}

\end{document}